# Quantifying the Online Long-Term Interest in Research

Murtuza Shahzad[*], Hamed Alhoori, Reva Freedman, Shaikh Abdul Rahman

**Abstract**

Research articles are being shared in increasing numbers on multiple online platforms. Although the scholarly impact of these articles has been widely studied, the online interest determined by how long the research articles are shared online remains unclear. Being cognizant of how long a research article is mentioned online could be valuable information to the researchers. In this paper, we analyzed multiple social media platforms on which users share and/or discuss scholarly articles. We built three clusters for papers, based on the number of yearly online mentions having publication dates ranging from the year 1920 to 2016. Using the online social media metrics for each of these three clusters, we built machine learning models to predict the long-term online interest in research articles. We addressed the prediction task with two different approaches: regression and classification. For the regression approach, the Multi-Layer Perceptron model performed best, and for the classification approach, the tree-based models performed better than other models. We found that old articles are most evident in the contexts of economics and industry (i.e., patents). In contrast, recently published articles are most evident in research platforms (i.e., Mendeley) followed by social media platforms (i.e., Twitter).

[*]Corresponding author
Email addresses: msyed1@niu.edu (Murtuza Shahzad), alhoori@niu.edu (Hamed Alhoori), freedman@cs.niu.edu (Reva Freedman), ashaikh2@niu.edu (Shaikh Abdul Rahman)

# 1. Introduction

Scholarly articles are being mentioned and shared online in increasing numbers and on many platforms, including scholar-focused platforms such as Mendeley and general platforms such as Twitter and Facebook. Online metrics about these research articles could be a valuable resource not only in determining trends in given research domains and subdomains but also in establishing how long discourse about these articles continues on social media platforms. If research articles have a long enough lifespan on social media platforms, the result may be that more and more online users get involved in discussing research interests, which could, in turn, lead to more research in a given domain. For example, at present, many people are sharing their research and opinions about machine learning and artificial intelligence on social media platforms. This extensive sharing has the potential to interest more researchers and students in the field and to increase the funding designated for that field. Further, accurate estimates of how long an article will have an online interest can be expected to be beneficial to understanding and measuring the societal impact of given research and the public's understanding of and interest in science.

Many research projects are executed in an ad hoc way inasmuch as they address a specific current problem and might have less interest in the long term. However, research that is sustainable over the long term has many benefits for all research stakeholders. It would be interesting, therefore, to determine how long people talk about any given research article on social media platforms. These social media metrics have emerged to be valuable metrics in measuring the impact of the research (Luc et al., 2021). It would be a worthwhile endeavor to determine how long discussions about given research content endure as a way to gauge public interest. As a lot of content related to research is shared online, discussions of research articles on social media could vary considerably from a matter of a few days to many years depending on numerous factors such as the domain of the research article, the online platforms on which it was shared, and the influence of the person or people who have shared the article. These online metrics could turn out to be valuable resources in estimating the online lifespan of research articles.

The majority of articles published may have a very limited lifespan, where the effort put into the research article is not rewarded, and, more importantly, the positive impact they could have is not realized online. It is pertinent to a researcher to know what social media platforms play an important role in disseminating the published scholarly work. Furthermore, the social media metrics would instigate the inquisitiveness in researchers to know for how long their work would be sustaining online. The main research question in this study is: *"How can we understand, analyze, and predict the online long-term interest in research articles"?*

Measuring the interest and impact of research through large-scale mining of scholarly data and altmetrics is still a largely unexplored area with many challenges and opportunities. There is a critical need to develop new approaches to confront these challenges and harness these opportunities by creating new metrics, building models, datasets, and software platforms that provide valuable insights into the use of scholarly literature and its impact within and beyond the scholarly community. In this paper, we propose to use this wealth of information about the social and media dissemination of research as an indicator of societal impact and to understand the online long-term interest in research, as it reflects not only immediate interest in a research finding but also the degree to which individuals find the work to be of sufficient interest to warrant online sharing after months or years of being published.

This study has several practical and theoretical implications. Our findings complement the literature of the science of science (Fortunato et al., 2018), sociology of science theories (Barnes et al., 1996; Merton, 1973), research policy (Bozeman, 2000), research impact (Penfield et al., 2014), and altmetrics (Sugimoto et al., 2017). Additionally, most literature on research evaluation has relied on citations analysis (Wouters et al., 2019). To the best of our knowledge, this is the first study that expands the idea of measuring the online long-term interest in research by introducing new features from social media platforms and measuring the online long-term interest. Thus, we extend the limited measurement of long-term research impact, or the impact of science on science, to a societal impact. Further, we were able to predict the lifespan of any given article using the features from multiple social media platforms. In addition, most previous studies relied on a limited number of papers, while this study has a large and diverse collection.

Further, the literature on altmetrics focused on a single or a few platforms over a limited number of years. Our study reveals the popularity of research across platforms over a long period of time. This can affect how researchers search for or share publications online. The findings are also beneficial to research stakeholders that are investigating broader research impact. Furthermore, the results would be valuable in designing new academic digital libraries and search engines by adding new features that would allow researchers to filter the literature based on the online long-term interest.

Given the potential usefulness of this research direction, we explored the trajectories of research articles on multiple social media platforms where users share research content. We also inspected how these platforms affect the online sustainability of the research articles. For this study, we considered a comprehensive timeline of research articles having publication dates ranging from 1920 through 2018. We analyzed the social media metrics for these articles and built machine learning models to predict how long a research article would last on social media. We examined this prediction task through the lens of both regression and classification approaches.

In summary, our contributions include: Proposing and evaluating the metric *Online Age* to measure and quantify the online interest in research articles; investigating the growth in online mentions of research articles on different online platforms for articles published from 1920 through 2018; developing machine learning models to predict the long-term online interest in research articles and identifying the most influential online sites that amplify this interest.

## 2. Related Work

*2.1. Obsolescence of Research Articles*

The literature includes multiple studies going back decades in which researchers have considered the life and obsolescence of scholarly articles by analyzing factors relevant to measuring the impact of scholarly research that is of interest to the public (Siravuri et al., 2018). Larivière et al. (2008) observed that the age of cited material has risen continuously since the mid-1960s. They observed that the citation life cycle of an article starts with a sudden increase in its initial years, followed by a peak, and finally by obsolescence i.e., the "decline over time in validity or utility of information" as defined by Line and Sandison (1974). Based on this definition of obsolescence as a relationship between use and time, the researchers proposed two kinds of literature studies in relation to obsolescence: (1) synchronous studies "made on records of use or bibliographic references made at one point in time, comparing the use against the age distribution of the material used or cited", and (2) diachronous studies "that follow the use of particular items through successive observations at different dates." Generally, obsolescence studies are synchronous because these are easier than diachronous studies to conduct. For functions of a continuous variable, Egghe (1994) defined the term rate of growth or obsolescence as an exponential function of the derivative of the log of the function. Egghe et al. (1995); Egghe (1993) observed that the rate of obsolescence varies and that this variation can be calculated as a utilization (mathematical) function. In the synchronous case, the larger the increase in production of research articles, the larger the obsolescence. In the diachronous (prospective) case, the opposite relation holds: the larger the increase in production, the lower the obsolescence rate. Stinson and Lancaster (1987) compared the synchronous and diachronous methods by analyzing both with respect to the dates of the publications referred to in 13,734 citations and 3,669 citations in diachronous and synchronous studies, respectively, over a 19-year period. They found that with the exception of the first two years, the approaches yielded similar statistical measures for the obsolescence of articles.

The concept of obsolescence has been applied to literature in various fields (Boxenbaum and Barnhill, 1984; Tsay, 2006). In a study of the obsolescence patterns of the U.S. geoscience literature, Kohut (1974) found that traditional fields such as paleontology and geology have an obsolescence period of over 20 years and that fast-changing fields such as solid earth geophysics have lower obsolescence rates. Using the synchronous approach, Gupta (1998) studied the growth and obsolescence of literature in theoretical population genetics and found that a high rate of growth in the literature does not mean a high rate of obsolescence for that literature; that there may not be any relationship between the growth rate and the obsolescence rate; and that there may not be any relationship between the growth rate of literature and the half-life of that literature. Sangam (1999) studied the obsolescence of literature in the field of psychology and found that compared with a slower growth rate, a higher growth rate in the literature is associated with greater obsolescence and a longer half-life. Cunningham and Bocock (1995) studied the obsolescence rate of articles in computer science subfields (networks and operating systems) and found a high obsolescence rate based on the median citation rate over a four-year period. Bouabid and Larivière (2013) found that the life expectancy of papers published in developed countries is on average shorter than that of papers published in emerging countries.

*2.2. Citation as Impact Indicator*

A standard metric used to measure scholarly impact is cited half-life, which is defined as "measure of citation survival measuring the number of years, going back from the current year, that covers 50% of the citations in the current year of the journal" (Garfield, 2001). As a journal is being cited by more articles, much of them are citing older literature. Datta et al. (2016) studied the half-life of software engineering research topics, taking into account over 19,000 papers from software engineering publication venues from 1975 to 2010. Obtained via natural language processing to identify the topics covered in each paper, their results showed that some research topics have a cited half-life of nearly 15 years.

Various factors have contributed to the increasing number of scholarly citations and the growing impact of scholarly articles in the recent past (Stacey, 2020). Barnett and Fink (2008) found that the invention of the internet increased the average life (citation age) of academic citations by 6 to 8 months. According to Šember et al. (2017), in the context of evolving technologies and methodologies, old articles are gaining new attention as authors refer to them in order to describe this evolution. Martín-Martín et al. (2016) verified results published by Google Scholar showing an increase in the number of citations of old articles published during the period 1990 to 2013. They surmised that the recent increase in the number of citations of older articles could be attributed to technology. They also commented that as Google is the most powerful search engine and the most useful for scholarly purposes, Google Scholar has been a significant factor in this growth.

There may be many underlying reasons for the trend whereby old papers are increasingly being cited in new papers, among which may be archival value. Oppenheim and Renn (1978) selected the most frequently cited old physics and physical chemistry articles (published before 1930) to determine why they continue to be cited many years after their publication date. They found that 40% of the citations of these old papers could be attributed to historical reasons, but that 60% could be attributed to the old papers remaining relevant to current research directions. As an extension of this work, Ahmed et al. (2004) explored the reasons why a paper by Watson and Crick (2003) continued to be cited frequently. Drawing on topology derived from previous research, they concluded that the article had been cited so many times because the authors who included it in their papers considered it important to the history of the research direction of which it is a part, had drawn on it as important to their research, and/or had offered criticism of it. In a study of citations of papers published between 1900 and 2006, Wallace et al. (2009) found that the citation trends were observed during the wars because of changes in the number of papers published. The increase in the citedness of the most recently published papers is accounted for by the high number of references for each paper. Avramescu (1979) explained the increased citation frequency with respect to the normal exponential decay of older articles. Huntington et al. (2006) considered the subject, search approach, and type of journal as possible factors in determining the age of articles cited and found that the age of the articles varies depending on the journal.

Analyzing and predicting important publications, citations, and author co-citations have been an active area of research (Savov et al., 2020; Bu et al., 2020). Citation of research publications is an indicator of how scientific knowledge spreads (Abramo et al., 2020; Liang et al., 2020). Stegehuis et al. (2015) proposed quantile-based regression models to predict future citations. Their models performed best when two variables—impact factor and early citation counts—were used together instead of separately. Wang et al. (2013) built a mechanistic model for the citation dynamics of papers from various journals and disciplines. They found that all papers tend to follow the same universal temporal pattern. In a study about the short and long-term citation windows, Wang and Zhang (2020) stated that the normalized citation indicator may not be reliable when a short citation window is used. To overcome this, they introduced a weighting factor using a correlation coefficient between citation counts of papers in the short citation window and in the fixed long citation window. The weight reflects the degree of reliability of the normalized citation indicator. Stern (2014) found that an article's ranking can be determined by initial citations of it, which can, in turn, determine the article's future citations. Sikdar et al. (2017) developed a concept for a reviewer–reviewer interaction network by studying papers from the Journal of High Energy Physics between 1997 and 2015. In the network, they considered features such as degree, clustering coefficient, closeness centrality, and betweenness centrality, all of which turned out to be strong predictors of long-term citations. Singh et al. (2017) found a negative correlation between early citations by high-impact authors and long-term citation count. Using linear regression models on Web of Science publications, Abramo et al. (2019) tested if the combination of a publication's early citations and the impact factor of the hosting journal could yield better prediction results for long-term citation counts. They found that the importance of a Journal's impact factor in the combination turns out to be insignificant after two years of publication.

*2.3. Online Impact of Scholarly Articles*

Numerous strategies for filtering scholarly work and assessing the impact of research have been developed (Alhoori et al., 2018). Peer review (Bornmann, 2011), citation analysis (Moed, 2005; Azoulay et al., 2018), and article-level assessment (McKeown et al., 2016) can all provide useful information about the impact of scientific research in this area, but these established methods have limitations and drawbacks (MacRoberts and MacRoberts, 1989, 1996; Seglen, 1997, 1992; Lima et al., 2013). They are time-consuming and self-limiting in that they exclude a large number of other channels for research attention and do not account for the holistic impact of scholarly outcomes (Piwowar, 2013; Priem and Hemminger, 2010; Bornmann, 2014). Members of the research community have argued that evaluating a scholarly article's impact solely on the basis of one metric is unlikely to provide an accurate picture of its value in this regard (Lima et al., 2013). Alternative measures have been proposed in recent years. For example, Neylon and Wu (2009) discovered that a variety of usage-based metrics, such as downloads, comments, and bookmarks, can be used to assess the impact of articles and journals, with each metric having distinct advantages and disadvantages.

Altmetrics have been proposed as a way to address some of the identified gaps (Priem et al., 2012). Altmetrics, which is gaining traction in the research community, refers to article-level metrics that have been proposed as a substitute for or supplement to traditional metrics. The critical distinction between traditional metrics (e.g., citations) and altmetrics is that, while the former quantify the impact of research within scholarly boundaries, the latter quantifies a variety of influences both within and beyond those boundaries. Altmetrics refers to a variety of Internet venues where scholarly works are referenced, stored, and/or shared (Das and Mishra, 2014; Melero, 2015; Chavda and Patel, 2016). Additionally, altmetrics can be used to quantify the impact of other scholarly products, including datasets, software, and presentations. As a result, a growing number of digital libraries and publishers are now including altmetrics on their websites. Altmetrics is concerned with quantifying not only societal impact but also scholarly impact. Recently, researchers were able to predict scholarly citations using altmetrics (Akella et al., 2021).

The impact of research on society is increasingly being considered by research communities (Samuel and Derrick, 2015; Bornmann, 2013; Shaikh and Alhoori, 2019). Additionally, a growing amount of scholarly content is shared and discussed on social media platforms on a daily basis (Ding et al., 2009; Fausto et al., 2012; Freeman et al., 2020, 2019). The number of research articles shared on these platforms is estimated to be increasing at a rate of 5–10% per month (Adie and Roe, 2013). In general, these platforms enable researchers to stay current on developments in their fields, as well as share and discuss their research data and findings, as well as solicit early feedback (Shahzad and Alhoori, 2022). Researchers include links to these news stories on their websites as evidence of their work's social impact. Tonia et al. (2016) found that exposure to social media did not have a significant effect on the articles' impact metrics such as downloads and citation counts. However, Allen et al. (2013) found that sharing articles in the clinical pain sciences on social media platforms such as Facebook, Twitter, LinkedIn, and ResearchBlogging.org led to an increase in the number of people who viewed and/or downloaded the articles. Using social media data to predict future citation counts, Thelwall and Nevill (2018) found that Mendeley's readership is an important predictive factor. Mohammadi et al. (2020) found that Facebook scholarly mentions are not very useful for predicting citations.

In summary, researchers have studied factors that contribute to the obsolescence, aging, and citation age of scholarly research. The invention of the internet and the development of scholarly digital libraries, search engines, and academic social platforms have increased the average lifespan of academic citations. As a result, old articles with archival value and/or that are important to current research have started to receive more citations. The citation life cycle of an article usually starts with a sudden increase in its initial years, followed by a peak, and finally by obsolescence. Studies of obsolescence have been conducted in relation to many research fields, and researchers have provided metrics to measure obsolescence. The study of citations reveals that articles reach their peak in relation to the number of citations accrued within two years of publication and then a gradual decrease in the number of citations accrued takes place thereafter. Many researchers have built models to predict the long-term citations of scholarly research. However, our study is the first to predict how long a given article is likely to be mentioned online. We predict the online lifespan of any given article using multiple social media platforms.

**3. Data**

We used a dataset from altmetric.com released in June 2018 and consisting of online mentions of about 19 million publications. The initial data were in JSON format, which we converted into a CSV file in order to perform model-building tasks. Of the data, which comprised 19,406,418 records, we considered only research articles published between 1920 and 2018, which reduced the dataset to 12,657,619 records (dataset A). The principal features included in our research are provided in Table 1. The features in Table 2 relate to the time of publication and the locations and positions (earliest or latest) of the online mentions.

All the temporal features described in Table 2 were used to generate two new features called *First Online Mention* and *Last Online Mention*, which provide information about the first and last online mention dates across all the platforms. We created a new feature, **Online Age**, which is the difference (number of months) between the date of the *First Online Mention* and the date of the *Last Online Mention*. This feature provides the number of months the article remains of interest to the online social media community.

We analyzed the articles in dataset A for three main reasons:

1. To observe the growth in the number of papers published from 1920 to 2018. The middle line (red color) in Figure 1 shows this growth on a log

Table 1: Descriptions of main features in the dataset.

| Feature | Description |
| --- | --- |
| Mendeley | Total number of mentions of a research article on Mendeley |
| CiteULike | Total number of mentions of a research article on CiteULike |
| Connotea | Total number of mentions of a research article on Connotea |
| Twitter | Total number of mentions of a research article on Twitter |
| Patent | Total number of mentions of a research article in patents |
| Facebook | Total number of mentions of a research article on Facebook |
| Blogs | Total number of mentions of a research article in blogs |
| Wikipedia | Total number of mentions of a research article on Wikipedia |
| Stack Overflow | Total number of mentions of a research article on Stack Overflow |
| Syllabi | Total number of mentions of a research article in syllabi |
| Policy | Total number of mentions of a research article in policy documents |
| News | Total number of mentions of a research article in news items |
| Google+ | Total number of mentions of a research article on Google+ |
| F1000 | Total number of mentions of a research article on F1000 |
| Reddit | Total number of mentions of a research article on Reddit |
| Video | Total number of mentions of a research article in online videos |
| Pinterest | Total number of mentions of a research article on Pinterest |
| Peer Review | Total number of mentions of a research article in peer reviews |
| Weibo | Total number of mentions of a research article on Weibo |
| LinkedIn | Total number of mentions of a research article on LinkedIn |
| Miscellaneous | Total number of miscellaneous online mentions |

scale.

2. To determine which articles to use in building our models. Of the articles

Table 2: Temporal features.

| Feature | Description |
| --- | --- |
| Publication date | Publication date of a research article |
| Twitter dates | First and last mention of a research article on Twitter |
| Patent dates | First and last mention of a research article in a patent |
| Facebook dates | First and last mention of a research article on Facebook |
| Blog dates | First and last mention of a research article in blogs |
| Wikipedia dates | First and last mention of a research article on Wikipedia |
| Stack Overflow dates | First and last mention of a research article on Stack Overflow |
| Syllabi dates | First and last mention of a research article in course syllabi (online courses with references to research) |
| Policy dates | First and last mention of a research article in policy documents |
| News dates | First and last mention of a research article on news outlets |
| First Online Mention | First online mention date of a research article across all the platforms |
| Last Online Mention | Last online mention date of a research article across all the platforms |

in dataset A, 8,520,926 articles (dataset B) have online mention dates, as shown on the log scale with the lower line (orange color) in Figure 1. The remaining articles had no information about online mention dates in the altmetric dataset.

3. To observe the number of online mentions of articles across the years. One research article can have multiple online mentions. Therefore, we plotted the number of online mentions of articles across the years, represented by the upper line (blue color) in Figure 1.

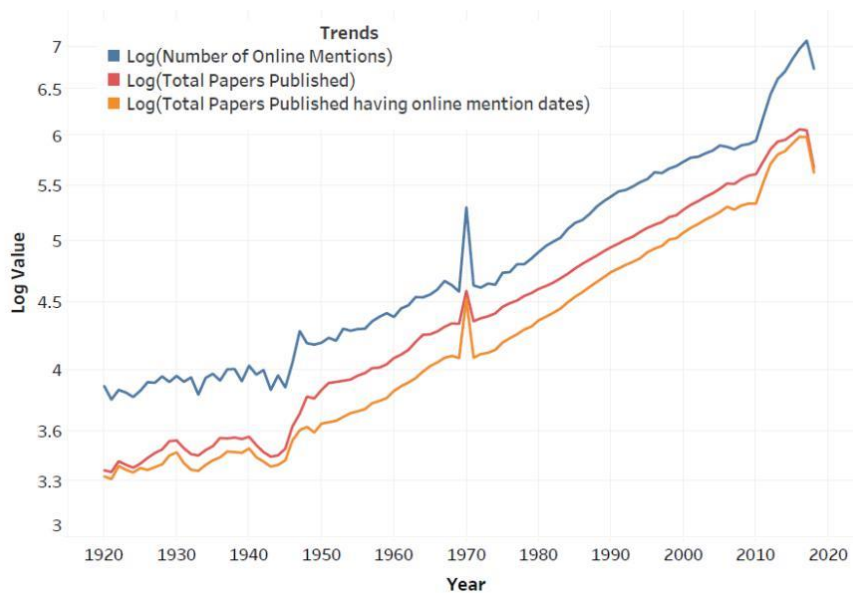

Figure 1: Growth of papers published and online mentions.

In Figure 1, the orange line below the red line shows that the dataset has some missing online dates. We observed a yearly increase in the number of articles mentioned online. We also observed that there is a plummet for the year 2018, as the altmetrics dataset used for the study had online mentions up until June 2018. Figure 2 shows the number of online mentions of research articles on various online platforms normalized using min-max scaling for all the platforms in each publication year. We observed that in comparison to all the other platforms, Mendeley accounts for the largest number of mentions in all years. Syllabi account for a larger proportion of the mentions in the years 1920 to 1970 than in the other years. Patent mentions have the second largest portion of mentions from 1970 to 2010 after Mendeley, and Twitter mentions account for an increasing proportion from 2010 to 2018.

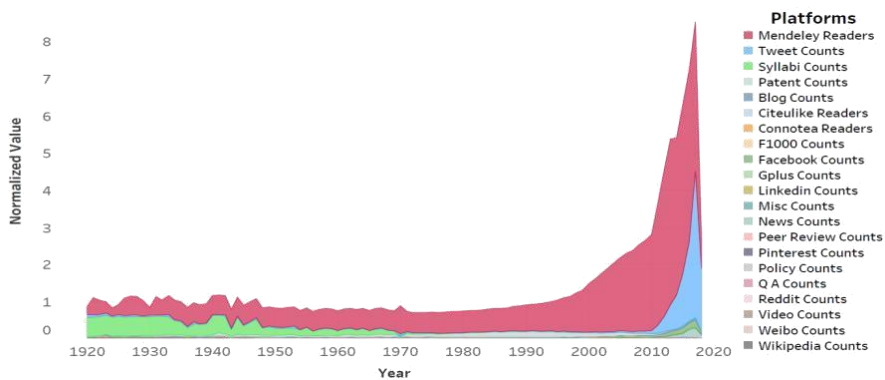

Figure 2: Online mentions on multiple platforms across publication years.

## 4. Methods

An active article on social media could be an article that has been consistently mentioned for an extended period of time. We tried several combinations for the number of platforms (e.g., at least one, two, or three platforms) and the frequency of online mention (e.g., every year or once every two years). We found that the Online Age for some of the previous combinations was very low or even zero in some cases. Therefore, we decided to define an **active article** as an article that was mentioned every year on at least three platforms since its first online mention up to 2018. On dataset B, we found that 242,164 articles satisfied the previous criteria for being considered as *active articles*. Most of the altmetrics for a research article are generally accrued around the time of the research article publication. For the articles published in the latest years in our dataset such as 2017 or 2018, the altmetrics may still be accumulating. Therefore, for the current dataset, we considered the articles with publication years up to 2016. For this curtailed dataset, we have 83,067 *active articles*.

Figure 3 summarizes our approach to measure the long-term online interest in research articles. At first, for understanding the long-term online interest in research articles, the natural inclination would be towards a time-series approach. However, the data that we have is not time-series data. In other words, we just have the altmetrics of research articles at one point in time (June 2018). Therefore, we eliminated the possibility of a time-series approach here. Next, we performed regression on the dataset (section 4.1). Based on the performance of the regression models, we observed that it is difficult for a model to learn from a dataset having a large variance in the publication years ranging from 1920 through 2016. As an alternative approach, we grouped the data into multiple clusters based on the publication years (section 4.2) and then applied regression on each cluster (section 4.2.1). We also considered the prediction of the long-term interest as a classification problem and built classifiers on all clusters (section 4.2.2).

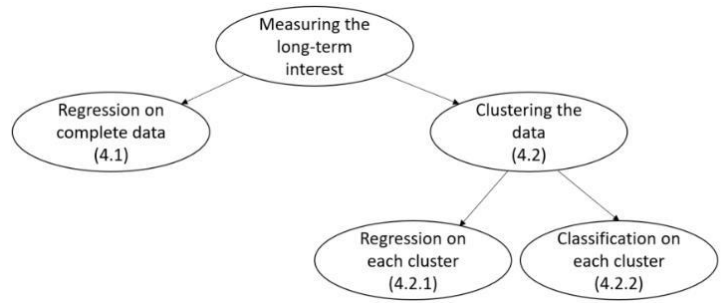

Figure 3: Our approach for measuring the long-term online interest in research articles.

*4.1. Regression on Complete Data*

To predict the long-term online interest in research articles, we need to predict how long a research paper remains on social media platforms. This could vary for different articles, from a few months to many years. To have uniformity for all of the research articles' online spans, we have considered the number of months a research article is mentioned on social media platforms, starting from its first online mention date. Using the features mentioned in Table 1, we build regression models on *active articles* to predict the *Online Age*. We split the dataset to 80/20 train-test ratio and used regression models from the scikit-learn implementation (Pedregosa et al., 2011). Table 3 shows the regression results for various models. After building some regression models, we observed that the models had low error measures but were poor in explaining the variance in the dependent variable. One possible reason for such results could be to account for the nature of the wider range of data with publication years from 1920 to 2016.

Table 3: Evaluation of Regression models.

| Model | Mean Absolute Error (MAE) | Root Mean Squared Error (RMSE) | R- Squared |
|---|---|---|---|
| Multiple Linear Regression | 18.30 | 25.01 | 0.29 |
| Decision Tree Regression | 14.49 | 21.78 | 0.46 |
| Random Forest Regression | 11.08 | 16.20 | 0.70 |

*4.2. Clustering the data*

As the regression models on the entire dataset were not able to yield satisfactory results in predicting the long-term online interest, we decided to treat the data points in the form of multiple clusters and then applied machine learning models to achieve better results. Clustering the data could be a better approach as it is unlikely for a single model to learn accurately from the data that spreads over almost a hundred years.

For clustering on dataset B, we used the elbow method (Thorndike, 1953) on the number of online mentions for each publication year. With this method, the k-means clustering algorithm performs clustering on various k-values to yield a plot for the variation of distances between the center of the cluster and each point for the different values of k. For a particular k, there would not be many variations in distance, and the plot is an elbow-like curve shape, which suggests the optimal number of clusters. In Figure 4, the optimal number of clusters is shown to be three. We, therefore, built three clusters using k-means on our data, as shown in Figure 5.

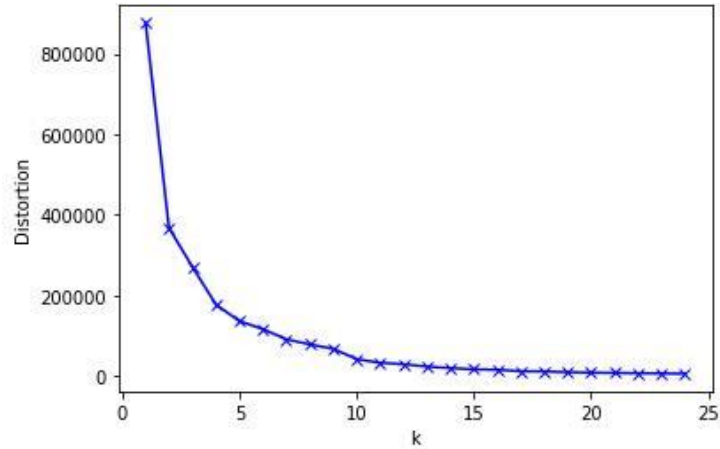

Figure 4: Elbow method to determine the optimal k-value.

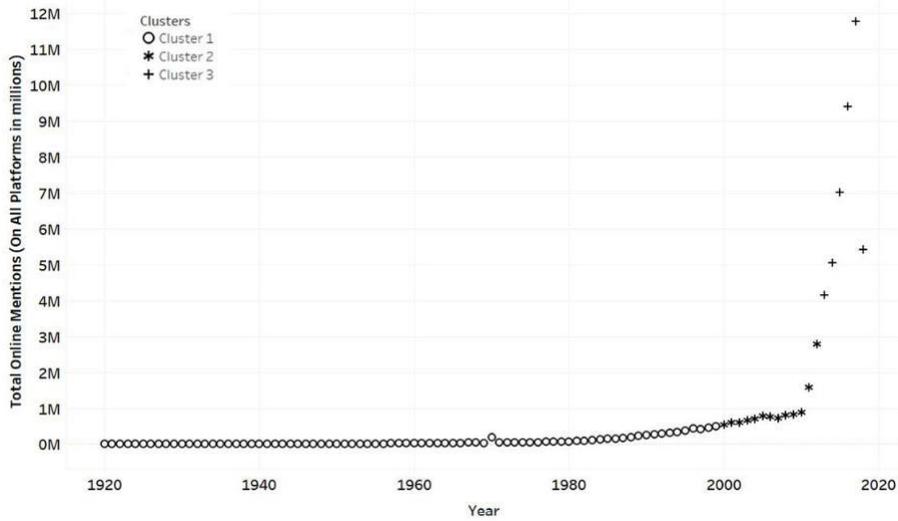

Figure 5: Three clusters using k-means.

Table 4 shows each cluster along with its details, such as the range of publication years, the total number of research articles in the cluster, and the total online mentions of all articles in that cluster. An increase in the number of records is evident from cluster 1 to cluster 3, which indicates that the number of mentions of newer research articles online increases over time.

Table 4: Details of the clusters formed.

| Cluster number | Range of publication years | Number of articles | Number of online mentions | Number of active articles |
|---|---|---|---|---|
| Cluster 1 | 1920–1999 | 1,538,350 | 6,588,889 | 4,641 |
| Cluster 2 | 2000–2012 | 2,699,017 | 12,272,590 | 27,077 |
| Cluster 3 | 2013–2016 | 2,998,724 | 25,552,821 | 51,349 |

*4.2.1. Regression on the Clustered Data*

In this approach, we treated the long-term interest as a regression problem and built machine learning models that predicted for how long a research paper remains on social media platforms. For the features mentioned in Table 1, we applied the scaling technique using scikit-learn's StandardScaler to normalize the data. We then split the data into 80/20 train-test ratio, built various regression models on all the clusters and predicted the *Online Age*. Additionally, we used Multi-layer Perceptron regressor to check if the results improve with Neural Networks. For the tree-based models, we applied 5-fold cross-validation with hyperparameter tuning to get the best parameters. We also present the feature importance for the Random Forest regression model to indicate the top 10 features in each cluster.

*4.2.2. Classification on the Clustered Data*

In this approach, we considered the long-term interest as a classification problem. We calculated the median of the *Online Age* feature for each cluster in terms of the number of months as shown in Table 5. We used the median as the main criterion, as it represents the center of the data and is not susceptible to outliers. We then built classification models with the features listed in Table1 for each of the three clusters in order to determine whether an article would receive an online mention that is equal to or greater than the median of the *Online Age*.

For classification, we trained and tested four algorithms using the scikit-learn implementation on our data: Random Forest, Decision Tree, Logistic Regression, and Gaussian Naive Bayes. For all the models, we performed 5-fold cross-validation with hyperparameter tuning to obtain better results. We also used the feature importance attribute (also known as Gini importance) of the Random Forest model and included the ten most important features in each cluster. We observed the performance of all clusters in terms of accuracy and weighted average scores for precision, recall, and F1.

Table 5: The median Online Age of the clusters.

| Cluster number | Range of publication years | Median Online Age (months) |
| --- | --- | --- |
| Cluster 1 | 1920–1999 | 16 |
| Cluster 2 | 2000–2012 | 34 |
| Cluster 3 | 2013–2016 | 25 |

## 5. Results

### 5.1. Regression Results

We built Multiple Linear Regression, Decision Tree Regression, Random Forest Regression, and Multi-layer Perceptron Regression models for each of the clusters. The performance of these regression models is shown in Table 6. We obtained the best parameters for tree-based models using hypermeter tuning.

### 5.1.1. Regression - cluster 1

The articles in cluster 1 have publication years of 1920 to 1999. The regression models built predicted the *Online Age* for this cluster. From Table 6, cluster 1, we can observe that tree-based models and Multi-Layer Perceptron gave lower error rates than the Multiple Linear Regression model. Random Forest and Multi-Layer Perceptron have higher $R^2$ values than other models indicating that these models better explain the variance of the dependent variable (*Online Age*). Figure 6 shows the ten most important features for the Random Forest model. We can notice that the Patent counts is the single most important feature with more than 80% of the feature importance share.

Table 6: The regression results of each cluster.

|  | Model | MAE | RMSE | $R^2$ |
|---|---|---|---|---|
| Cluster 1 | Multiple Linear Regression | 34.33 | 53.14 | 0.35 |
|  | Decision Tree | 16.93 | 31.61 | 0.77 |
|  | Random Forest | 14.24 | 23.98 | 0.86 |
|  | Multi-layer Perceptron | 15.20 | 24.98 | 0.85 |
| Cluster 2 | Multiple Linear Regression | 28.83 | 33.63 | 0.26 |
|  | Decision Tree | 13.29 | 21.72 | 0.69 |
|  | Random Forest | 10.12 | 15.99 | 0.83 |
|  | Multi-layer Perceptron | 10.11 | 15.55 | 0.84 |
| Cluster 3 | Multiple Linear Regression | 10.29 | 13.71 | 0.11 |
|  | Decision Tree | 9.70 | 12.72 | 0.24 |
|  | Random Forest | 8.66 | 11.22 | 0.40 |
|  | Multi-layer Perceptron | 8.18 | 10.62 | 0.47 |

*5.1.2. Regression - cluster 2*

The articles in cluster 2 have publication years of 2000 to 2012. We built regression models that predicted the *Online Age* for this cluster. From Table 6, cluster 2, we can see that Random Forest and Multi-Layer Perceptron models performed better with lower errors measures than Decision Tree and Multiple Linear Regression. Random Forest and Multi-Layer Perceptron model better explain the variance in the dependent variable by having an $R^2$ value of 0.83 and 0.84 respectively. Figure 7 shows the ten most important features for the Random Forest model. We can notice that the most important feature for the prediction is the Tweet counts followed by Patent counts. These two features constitute approximately 80% of the feature importance share.

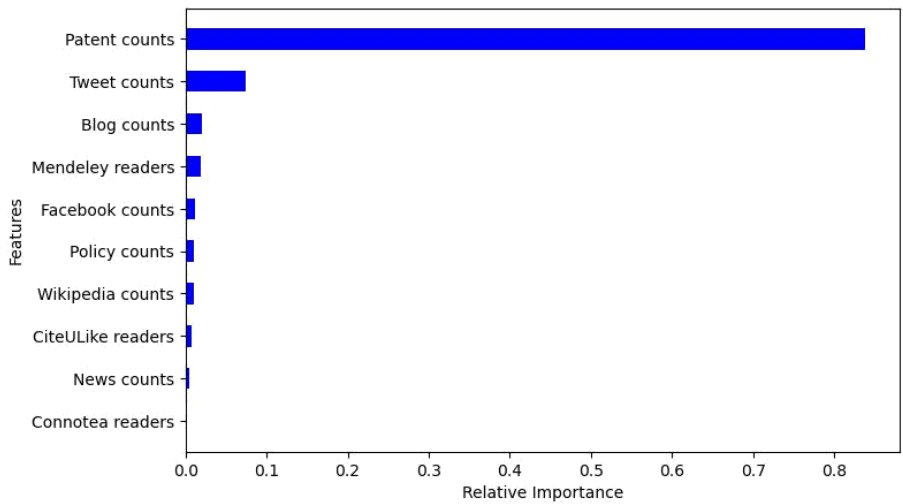

Figure 6: Regression - The ten most important features for cluster 1 (1920-1999).

*5.1.3. Regression - cluster 3*

Cluster 3 has articles that were published in the years 2013 through 2016. Table 6 shows the performance of the regression models to predict the *Online Age*. We observed that Random Forest and Multi-Layer Perceptron models have lower error rates and are better able to explain the variance of the dependent variable. Figure 8 shows the ten most important features for the Random Forest model. We found that Tweet count and Mendeley's readership are the most important features for the Random Forest model with approximately 40% of the feature importance share. On comparing the results of regression models in cluster 3 with clusters 1 and 2, we notice that cluster 3 has lower error measures but poorly explains the variance in the dependent variable.

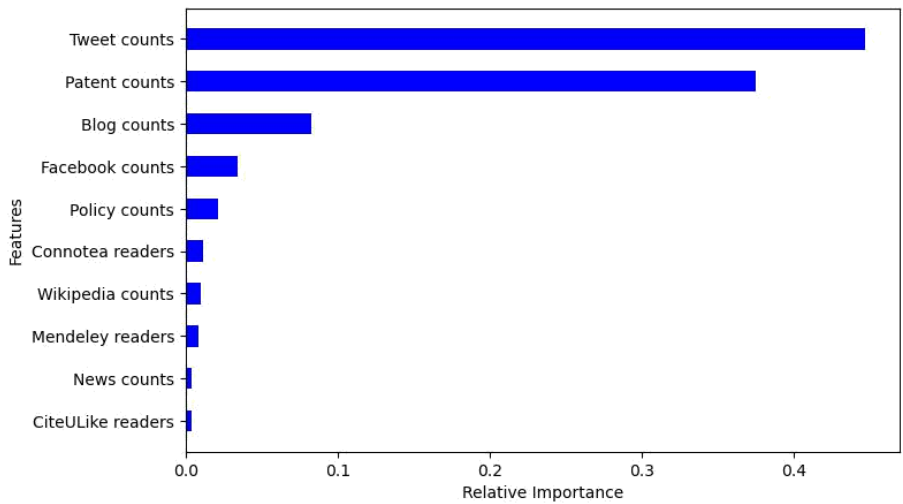

Figure 7: Regression - The ten most important features for cluster 2 (2000-2012).

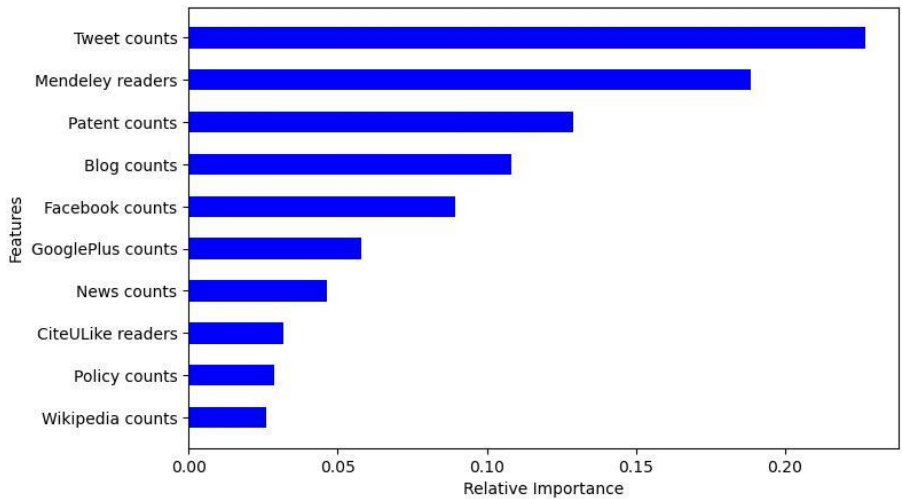

Figure 8: Regression - The ten most important features for cluster 3 (2013-2016).

## *5.2. Classification Results*

### *5.2.1. Classification - cluster 1*

Cluster 1 comprises articles published in the period of 1920 to 1999. The median *Online Age* for this cluster is 16 months. We built classification models to predict whether the *Online Age* would be greater than or equal to the median. Figure 9 shows the performance of the models. Achieving accuracy of 91%, the Random Forest model performed the best. Figure 10 shows the ten most important features for the Random Forest model. We observed that for cluster 1, the Patent count is the most important feature, followed by the Tweet count. These two features accounted for approximately 55% of the feature importance share. From Figure 2, we also noticed that for the years 1920–1970, the Syllabi count accounted for a large proportion of online share whereas for the years 1971–2000 Patent count accounted for a large proportion.

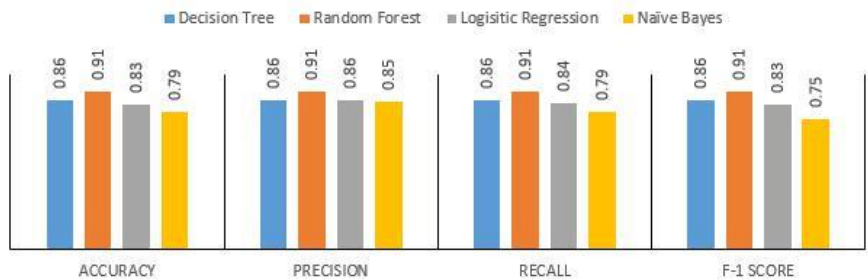

Figure 9: Classification - Performance of models in cluster 1 (1920-1999).

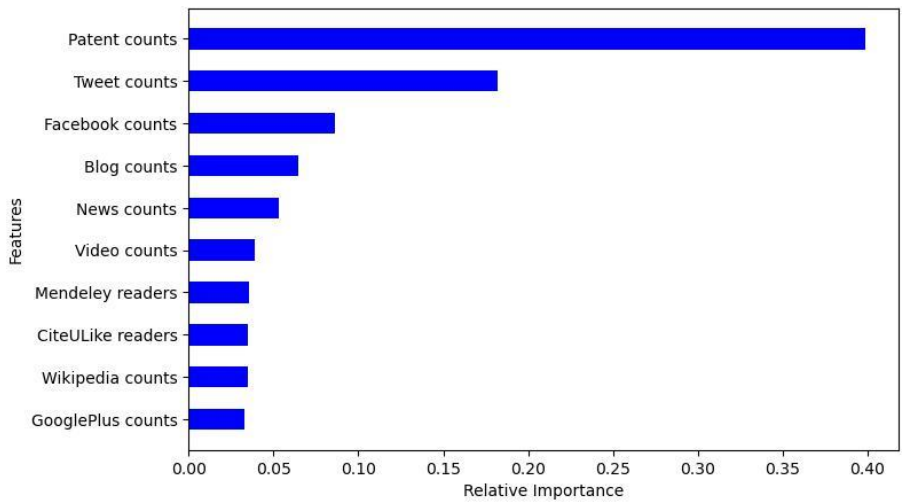

Figure 10: Classification - The ten most important features for cluster 1 (1920-1999).

*5.2.2. Classification - cluster 2*

The articles published in the period 2000 to 2012 fall under cluster 2. As shown in Table 5, the median *Online Age* for this cluster is 34 months. The Decision Tree and Random Forest classifiers performed better than the other classifiers, with all the performance metrics having a score of 92% and 94%, respectively for these two models, as shown in Figure 11. The ten most important features for the Random Forest model of this cluster are shown in Figure 12. We observed that the Patent count was the most important feature, accounting for about 35% of the feature importance share, followed by Twitter and Facebook counts.

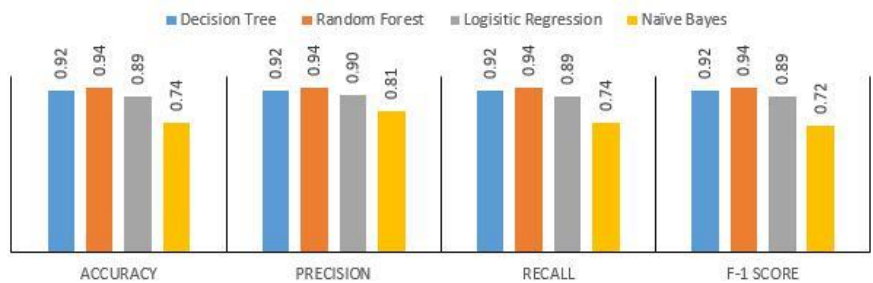

Figure 11: Classification - Performance of models in cluster 2 (2000-2012).

*5.2.3. Classification - cluster 3*

Cluster 3 consists of research articles published from 2013 to 2016. These articles have a median *Online Age* of 25 months. Figure 13 shows that with an accuracy of 73%, the Random Forest model performed better than the other models. Figure 14 shows the ten most important features for the Random Forest classifier: Mendeley's readership and Tweet count constitute about 40% of the feature importance share.

In addition to the performance metrics and feature importance, we plotted the Receiver Operating Characteristic (ROC) curve for all three clusters, as shown in Figure 15. We observed that for all 3 clusters the area under the curves is slightly more for tree-based classifiers and Logistic Regression in comparison with the

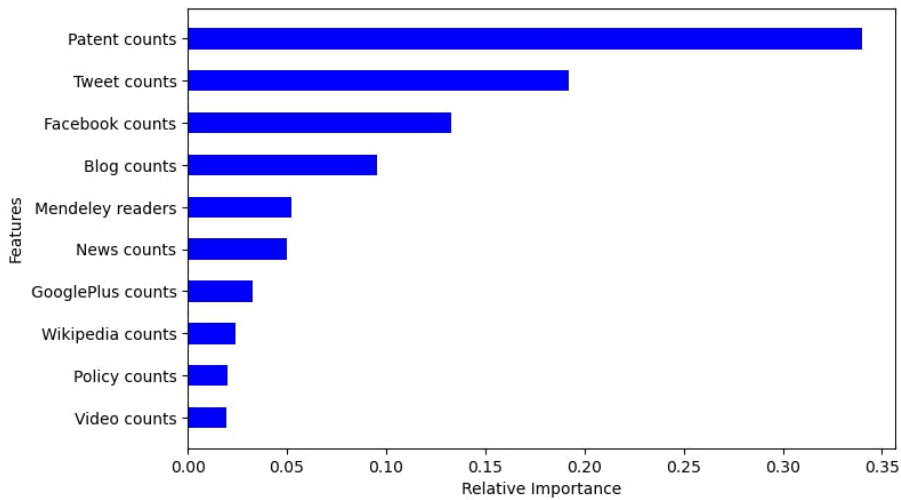

Figure 12: Classification - The ten most important features for cluster 2 (2000-2012).

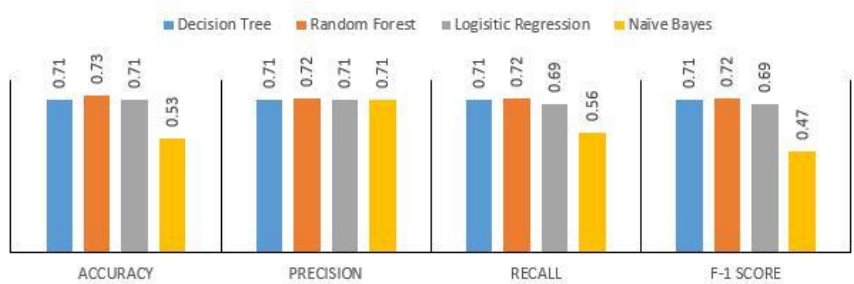

Figure 13: Classification - Performance of models in cluster 3 (2013-2016).

Naïve Bayes classifier.

## 6. Discussion

In this experiment designed to predict the long-term interest in research articles in terms of the number of months they last on online platforms after their first online mention, we used the counts of online mentions for research articles on multiple online platforms as the features for building machine learning models. We split the data pertaining to research papers published in the years 1920 to 2016 into 3 clusters based on the elbow methodology for k-means clustering.

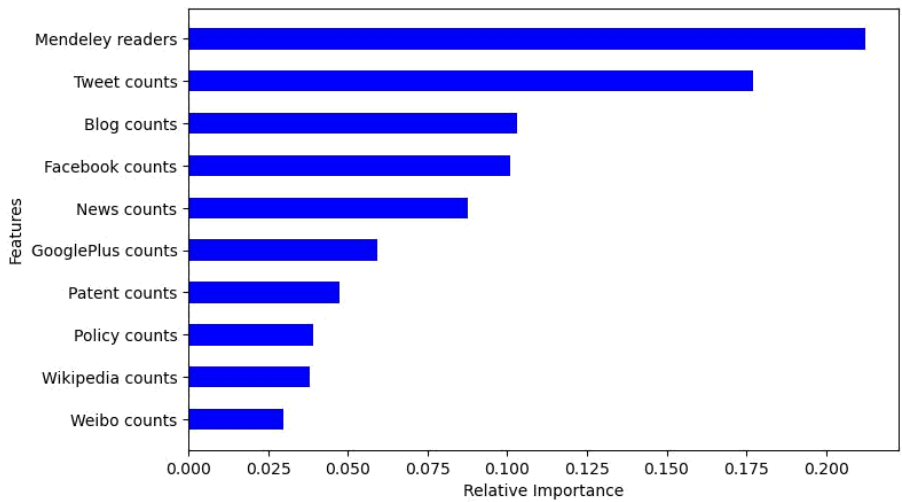

Figure 14: Classification - The ten most important features for cluster 3 (2013-2016).

For each cluster, we calculated the median *Online Age*, which served as a criterion for long-term interest. We built machine learning models with 5-fold cross-validation. For all three clusters, the Naive Bayes model performed worse than the other models for classification. We observed that the relative importance of the online platforms for prediction differs across the clusters. Table 7 shows a summary of the most important features in each cluster.

When investigating references to articles, the use of course syllabi is not a popular approach to evaluating research. However, we included this aspect in our investigation. In our dataset, older articles have more influence on education than do recently published articles. However, it is limited by our dataset, which in terms of syllabi includes only those available on the internet. Various studies have found that course syllabi could be useful to measure the teaching impact of publications, especially in the humanities and social sciences (Thelwall and Kousha, 2015; Kousha and Thelwall, 2016, 2008). Our dataset does not include the subjects of articles that are mentioned in syllabi derived from the Open Syllabus Project (OSP). Through a manual check on 20 random articles in the OSP, we found that several are related to humanities and social sciences. We found that older

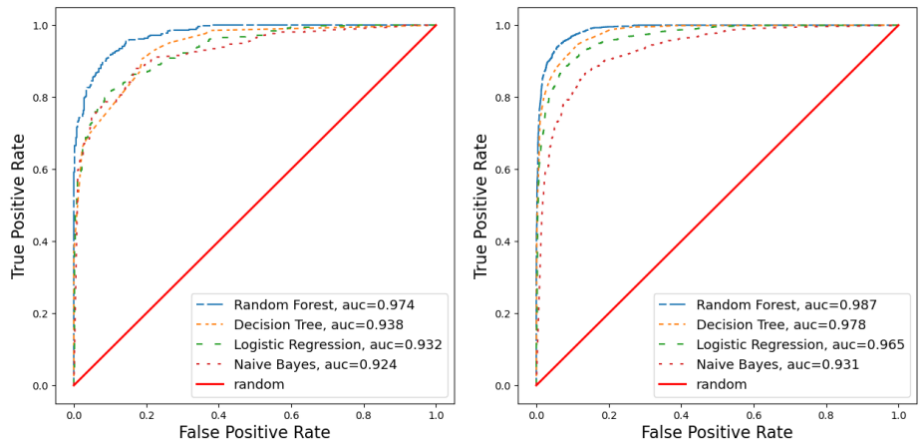

(a) ROC for cluster 1(1920-1999)

(b) ROC for cluster 2(2000-2012)

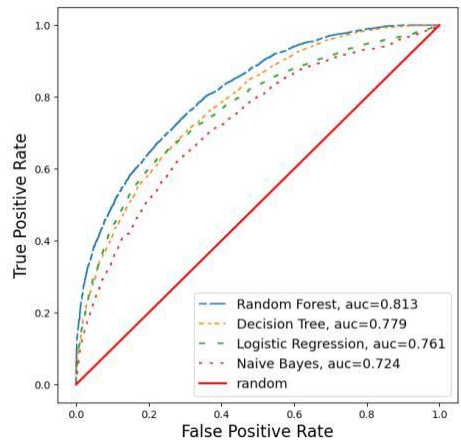

(c) ROC for cluster 3(2013-2016)

Figure 15: Receiver Operating Characteristic (ROC) curves for all clusters.

articles have more impact on patents, as is evident in cluster 2.

Table 7: Summary of important platforms for prediction in each cluster.

| Cluster | Publication years of research articles | Most important for prediction - Regression | Most important for prediction - Classification |
|---|---|---|---|
| 1 | 1920–1999 | Patent | Patent and Twitter |
| 2 | 2000–2012 | Twitter and Patent | Patent and Twitter |
| 3 | 2013–2016 | Twitter and Mendeley | Mendeley and Twitter |

In relation to cluster 3, we observed that Mendeley's readership and Twitter are important in predicting the online interest in research articles. For the regression models, we also noticed that the Multi-Layer Perceptron did perform better than the Random Forest model for cluster 2 and 3. We also observed that having a smaller cluster (in terms of the range of publication years) achieved lower errors in regression models, as can be seen in Table 6

In the current study, in order to measure the online long-term interest in research articles, we used the latest online mentions of articles on online platforms regardless of how many times those articles had been mentioned online during the focal period. Yet a high number of mentions for a new article could imply that it is popular, which could lead, in turn, to long-term interest even though our current study would not recognize this fact since such a paper would be too new to have a large half-life. This is a limitation of the current study, which we plan to address in future work by collecting a new dataset. Another limitation of this study is that the data that we have used is not time-series data. In the future, we plan to collect altmetrics data at several time intervals to predict the online long-term interest at any point in time.

## 7. Conclusion and Future Work

In this study, we created models to predict the online long-term interest in research articles on social media. We found that the number of mentions in Patents documents, Mendeley and Twitter are the main factors in determining the long-term online interest for an article. We observed that research articles used in patents are usually old published articles that have been studied extensively and proven to be valid and trustworthy. Further, articles published before a few years are more seen in online reference systems such as Mendeley. In addition, mentions of articles were more numerous on social media platforms such as Twitter within days to months of publication. We also observed that of all the models tested, the Decision Tree and Random Forest performed best in the classification approach, and Multi-Layer Perceptron performed best in the Regression approach. In future work, we plan to use a range of prediction categories such as short-, mid-, and long-term interest. Additionally, we intend to include more features such as the textual features in research articles, citation count, and the h-index of the authors and venues. Further, we will study the differences between journals, disciplines, and countries in regard to a research article's lifespan.

**Acknowledgments**

This work is supported in part by NSF Grant No. 2022443.